\documentclass[aps,prl,twocolumn,showpacs,superscriptaddress]{revtex4-1}
\usepackage{graphicx}
\usepackage{bm}
\usepackage{bbm}

\usepackage{subfigure}
\usepackage{amssymb}

\usepackage{xcolor,xspace,colortbl,rotating}
\usepackage{amsfonts}
\usepackage{tabulary}

\begin{document}
\title{Ultrafast amplification and non-linear magneto-elastic coupling of coherent magnon modes in an antiferromagnet}

\author{D. Bossini} \email[]{davide.bossini@uni-konstanz.de}
\affiliation{Department of Physics and Center for Applied Photonics, University of Konstanz, D-78457 Konstanz, Germany.}
\author{M. Pancaldi}
\affiliation{Department of Physics, Stockholm University, 106 91, Stockholm, Sweden}
\author{L. Soumah}
\affiliation{Department of Physics, Stockholm University, 106 91, Stockholm, Sweden}
\author{M. Basini}
\affiliation{Department of Physics, Stockholm University, 106 91, Stockholm, Sweden}
\author{F. Mertens}
\affiliation{Experimentelle Physik VI, 
Technische Universit\"at Dortmund, 
Otto-Hahn Stra\ss{}e 4, 44227 Dortmund, Germany}
\author{M. Cinchetti}
\affiliation{Experimentelle Physik VI, 
Technische Universit\"at Dortmund, 
Otto-Hahn Stra\ss{}e 4, 44227 Dortmund, Germany}
\author{T. Satoh}
\affiliation{Department of Physics, Tokyo Institute of Technology, Tokyo 152-8551, Japan}
\author{O. Gomonay}
\affiliation{Institut f\"{u}r Physik, Johannes Gutenberg Universit\"{a}t Mainz, D-55099 Mainz, Germany}
\author{S. Bonetti}

\affiliation{Department of Physics, Stockholm University, 106 91, Stockholm, Sweden}
\affiliation{Department of Molecular Sciences and Nanosystems, Ca' Foscari University of Venice, 30172 Venezia-Mestre, Italy.}

\begin{abstract}
We investigate the role of domain walls in the ultrafast magnon dynamics of an antiferromagnetic NiO single crystal in a pump-probe experiment with variable pump photon energy. Analysing the amplitude of the energy-dependent photo-induced ultrafast spin dynamics, we detect a yet unreported coupling between the material's characteristic THz- and a GHz-magnon modes. We explain this unexpected coupling between two orthogonal eigenstates of the corresponding Hamiltonian by modelling the magneto-elastic interaction between spins in different domains. We find that such interaction, in the non-linear regime, couples the two different magnon modes via the domain walls and it can be optically exploited via the exciton-magnon resonance.
\end{abstract}

\maketitle

\pagenumbering{arabic}

Antiferromagnets (AFs) have recently surged as candidates for a novel paradigm of spintronics devices able to outperform ferro- and ferrimagnetic materials in terms of operational frequency, storage density and resilience to external fields\cite{Wadley2016a,Lebrun:2018kt,Baltz:2018iv,Gomonay:2018js}. Intrinsically the long-range antiferromagnetic order presents domains, which can hardly be manipulated. This magnetic texture and the magneto-elastic coupling - which is intimately interconnected to the domain structure - have been very recently shown to play a major role in the mechanism allowing electric manipulations of the N\'eel vector\cite{Baldrati2020,Meer2020,Zhang:2019gc}. The quest for an ever faster and more energy efficient control of AFs motivates the use of ultrashort light pulses as stimulus to drive (sub)-picosecond spin dynamics\cite{Nemec:2018gw,Satoh2010,Baierl:2016cb,Kampfrath2010,Bossini2016,Tzschaschel:2017eh,Nishitani2010,Simoncig:2017bn,Kanda:2011ja,Bossini:2019in}. However, the role of domain walls in magneto-elastic AFs on the ultrafast N\'eel vector dynamics has been hitherto not addressed, although being a crucial issue, since the overwhelming majority of AFs in nature display a multidomain magnetoelastic groundstate.

Here we demonstrate that the domain walls can activate a novel functionality in an antiferromagnetic crystal, namely a non-linear magneto-elastic domain-walls-mediated coupling between coherent spin wave modes belonging to different branches of the magnon dispersion, affecting the ultrafast dynamics of the N\'eel vector. We realise experimentally the tailored amplification of coherent THz oscillations of the N\'eel vector by pumping a magnon mode in an antiferromagnetic NiO crystal. This process is triggered by driving a combined electronic and magnetic transition and results even in the amplification of a different GHz magnon mode via the aforementioned coupling. Finally, we formulate a macroscopic phenomenological model able to explain the observations by taking into account the role of the domain walls in the ultrafast dynamics of the N\'eel vector.

\begin{figure}[t]
	\centering
	\includegraphics[width=\columnwidth]{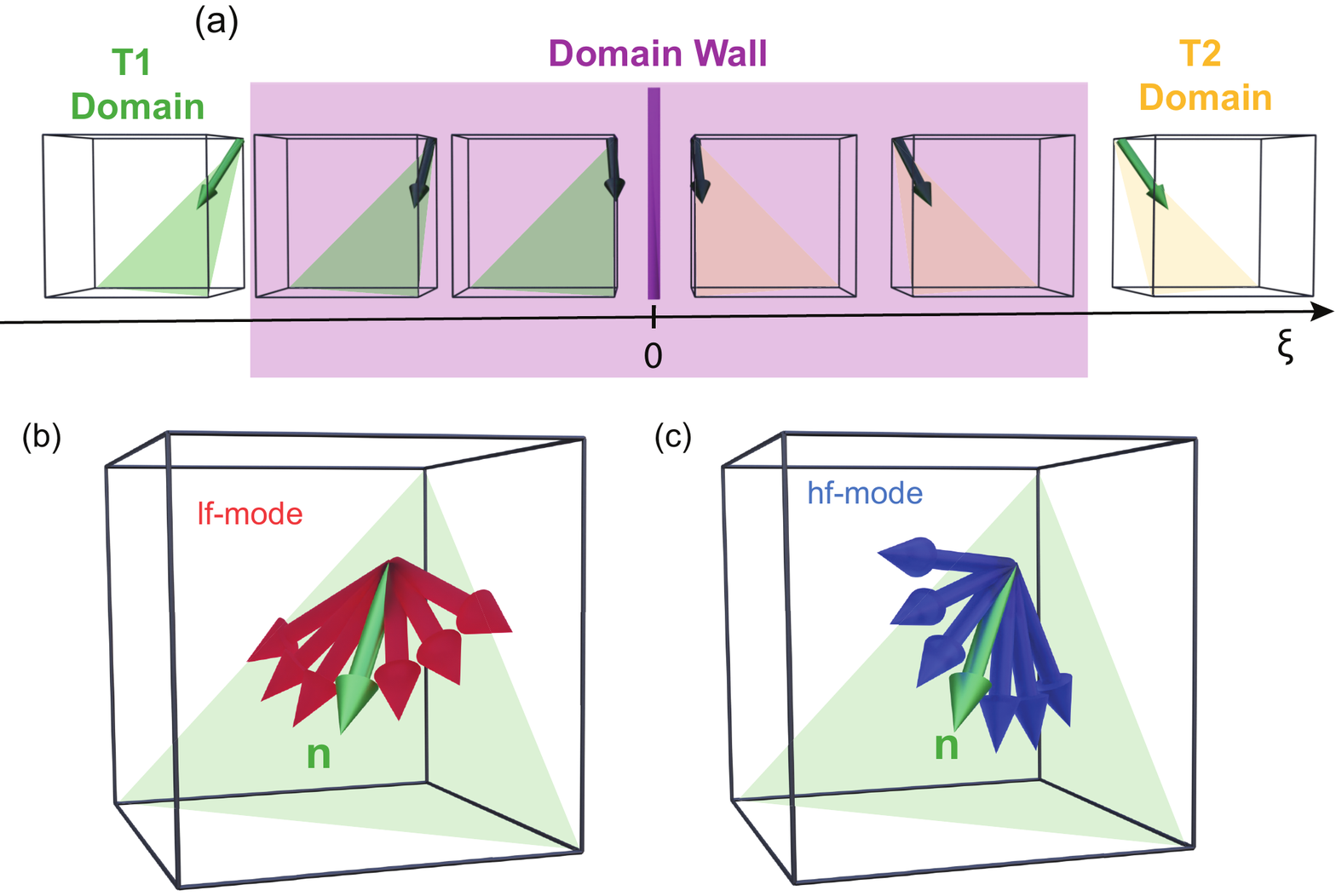}
	\caption{\footnotesize{(a) Two of the four possible T domains and the corresponding orientation of the N\'eel vector (green arrows). The magenta area represents the wall; the order parameter (dark arrows) rotates in this region. At the center of the wall ($\xi \approx 0$) the orientations of $\mathbf{n}$ in the T1 and T2 domain are parallel with each other. (b) In- and (c) out-of-plane dynamics of the order parameter induced by the low- and high-frequency magnon mode.}}
	\label{fig:Fig2}
	\end{figure}

Our specimen is a 100 $\mu$m-thick free-standing single crystal of NiO, cut along the $\langle 111 \rangle$ direction and has a multidomain structure. A specimen in a multidomain state can be described invoking as many antiferromagnetic vectors (defined as $\mathbf{n} \equiv \mathbf{M}^{\Uparrow}- \mathbf{M}^{\Downarrow}$, where $\mathbf{M}^{\Uparrow,\Downarrow}$ represent the magnetisation of the two sublattices.), each belonging to a T-domain (Fig. \ref{fig:Fig2}(a)). 

The domain structure of NiO, comprising spin (S) and twin (T) domains, is tightly connected with the magneto-elastic coupling, since when the crystal enters the magnetic phase strained magnetic domains are formed~\cite{Baruchel:1977hy,Arai2012}. The magneto-elastic energy is also the major contribution to the the anisotropy gap in the magnon dispersion~\cite{Ozhogin:1976hu} .

	\begin{figure}[t]
	\centering
	\includegraphics[width=\columnwidth]{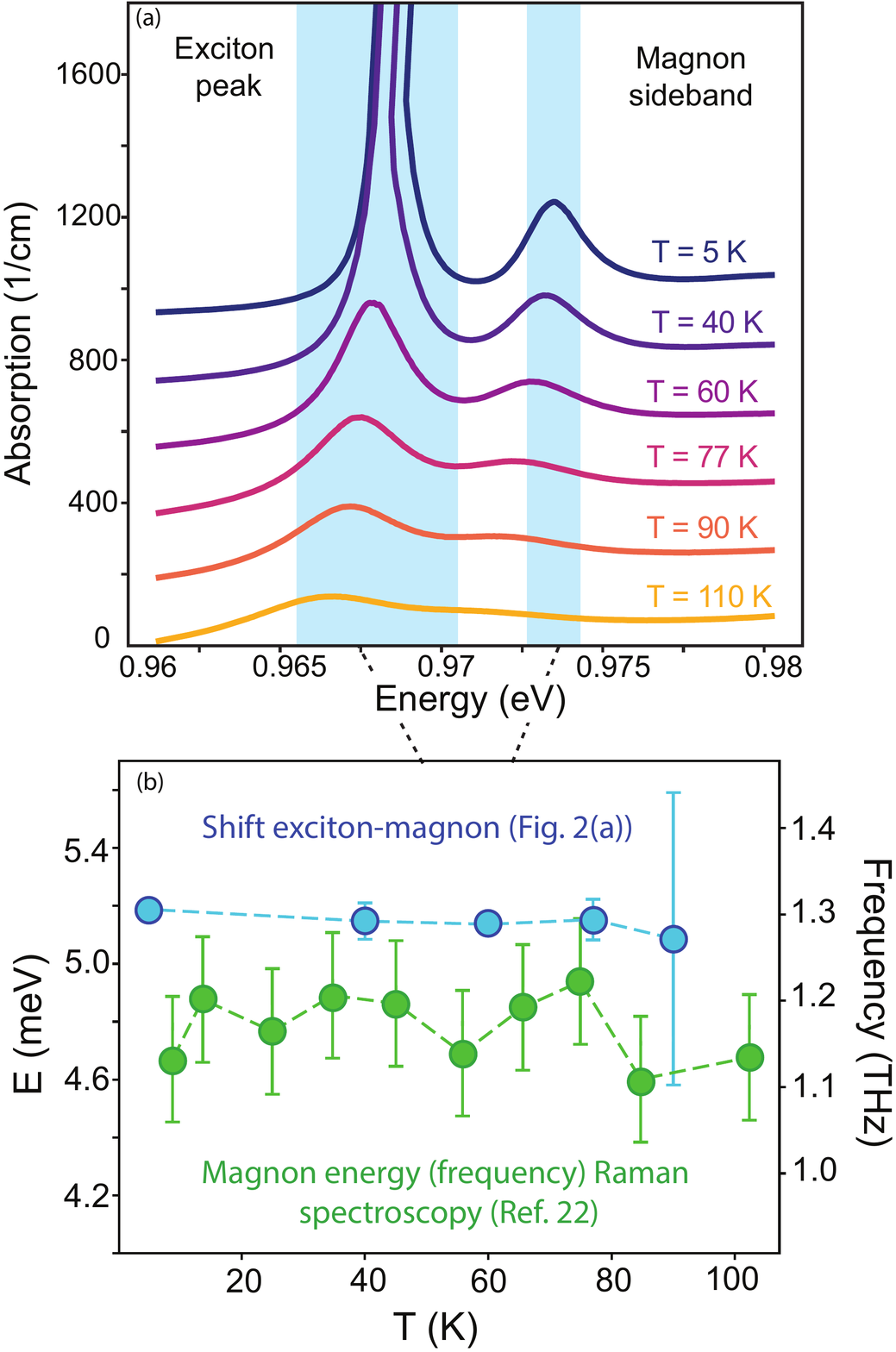}
	\caption{\footnotesize{(a) Absorption coefficient of NiO around the exciton-magnon resonance detected at different temperatures. Traces are displaced vertically for the sake of clarity. (b) Blue symbols: temperature-dependent energy shift between the electronic peak and the sideband~\cite{SM}. Green symbols: reported values of the temperature dependence of the 1 THz magnon mode, detected with Raman spectroscopy~\cite{grimsditch1998}}.}
	\label{fig:Fig3}
	\end{figure}
	
As a matter of fact, the magnon dispersion of NiO can be described in the first approximation in terms of two branches, so that at the center of the Brillouin zone two modes are active (Fig. \ref{fig:Fig2}(b)-(c)). A 1.07 THz mode, which we are going to refer to as the \textit{high-frequency} (hf) mode, has already been excited by means of two different approaches: resonant THz excitation\cite{Kampfrath2010,Baierl:2016cb} and non-resonant impulsive stimulated Raman scattering~\cite{Satoh2010,Tzschaschel:2017eh,Simoncig:2017bn,Kanda:2011ja,Takahara:2012cp}(ISRS) mainly by means of optical laser pulses. The ISRS mechanism succeeded also in inducing a lower frequency magnon mode, which will be labeled as \textit{low-frequency} (lf), with a frequency on the order of 130 GHz\cite{Satoh2010,Tzschaschel:2017eh}. The previous time-resolved investigations reporting the photo-activation of both modes, mostly performed focussing the pump and probe beams into a single T-domain, do not show any form of coupling or interaction between the two magnetic eigenmodes. 
 
An unexplored pathway to the femtosecond optical generation of the hf-mode relies on the \textit{exciton-magnon} (X-M) transition\cite{Tanabe1965,Tanabe:1982vb}. This process consists in the simultaneous excitation of a spin-forbidden (i.e. $\Delta  S$ = 1) electronic transition and of a magnon (i.e. $\Delta  S$ = -1), restoring the overall conservation of spin as required for electric dipole transitions\cite{Tanabe1965,Tanabe:1982vb}. We thus measured the absorption spectrum of our sample as a function of temperature (for the details see~\cite{SM}). The spectra obtained for $T$ < 100 K  display a peak centred at approximately 0.97 eV and a sideband at higher energy (Fig. \ref{fig:Fig3}(a)). The position of the sideband is temperature dependent, and the energy shift between the two spectral features ($\approx$ 4 meV) is consistent with the energy of the hf-mode (Fig. \ref{fig:Fig3}(b)) observed by Raman spectroscopy~\cite{grimsditch1998}. Our observations are in excellent agreement with the literature\cite{MironovaUlmane:2012ca,grimsditch1998}, so resonantly pumping our sample in the 0.97 eV spectral range is expected to result in the generation of the hf-mode. We aim at answering two open scientific questions. First, whether the X-M transition can actually resonantly induce coherent magnons on the femtosecond time-scale. Second, whether the domain walls play a role in the ultrafast spin dynamics of a multidomain AF and, in case they do, what this role is.

	\begin{figure}[t]
	\centering
	\includegraphics[width=\columnwidth]{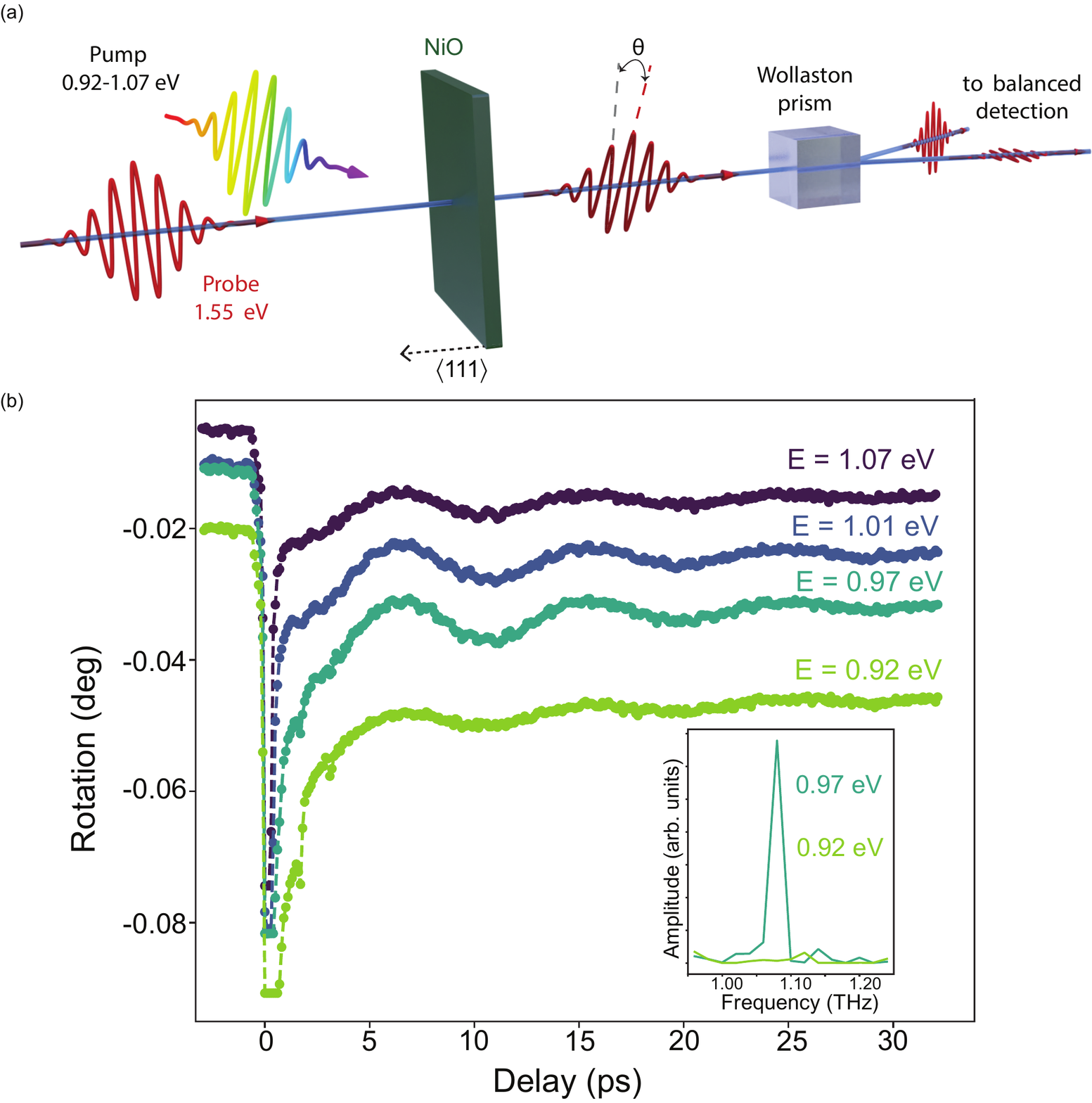}
	\caption{\footnotesize{(a) Schematic representation of the set-up. (b) Selected pump-probe traces for different pump-photon energies. The probe photon energy is 1.55 eV in each data set. The temperature of the sample was set to 77 K. The polarization of the pump beam is linear and parallel to the $[11\overline{2}]$ direction, while the probe beam was linearly polarized  45$^{\circ}$ away from the $[11\overline{2}]$ axis. The pump fluence was kept constant to $\approx$ 10 mJ/cm$^2$. We have not performed fluence-dependent measurements, as increasing the fluence results in damaging the sample and lower fluence value implied an unfavourable signal/noise ratio. The incoherent background is ascribed to heating of both the lattice and the magnetic system, consistently with the literature~\cite{Bossini2014,Kimel2002}.}}
	\label{fig:Fig4}
	\end{figure}

We tackle these questions in a magneto-optical pump-probe experiment, in which the pump photon energy can be tuned in the 0.92 - 1.07 eV spectral range (Fig. \ref{fig:Fig4}(a)), allowing to compare the spin dynamics triggered by a resonant pumping of the X-M with the signal detected by exciting NiO non-resonantly (set-up described in~\cite{SM}). The rotation of the polarization detected in every trace shown in Fig. \ref{fig:Fig4}(b) reveal oscillations at the frequency of approximately 110 GHz. Considering the value of the frequency, we ascribe this harmonic component of the signal to the lf-mode\cite{Satoh2010,Tzschaschel:2017eh}. The slight deviation of the frequency from the reported value is due to magnetostriction induced by both internal and external strains of the sample, which can significantly affect the magnon frequency in antiferromagnets\cite{Ganot:1982bj}.  Additionally, some of the traces in Fig. \ref{fig:Fig4}(b) display also a faster oscillatory component, whose frequency matches the reported 1.07 THz value of the hf-mode. In the inset of Fig. \ref{fig:Fig4}(b) the power spectra (i.e. square modulus of the Fourier transform) of the 0.92 eV and 0.97 eV time-traces are shown, displaying the presence and absence of the 1 THz magnon. The discussion of the magneto-optical effects involved in our experiments is reported in~\cite{SM}.

\begin{figure}[t]
	\centering
	\includegraphics[width = \columnwidth]{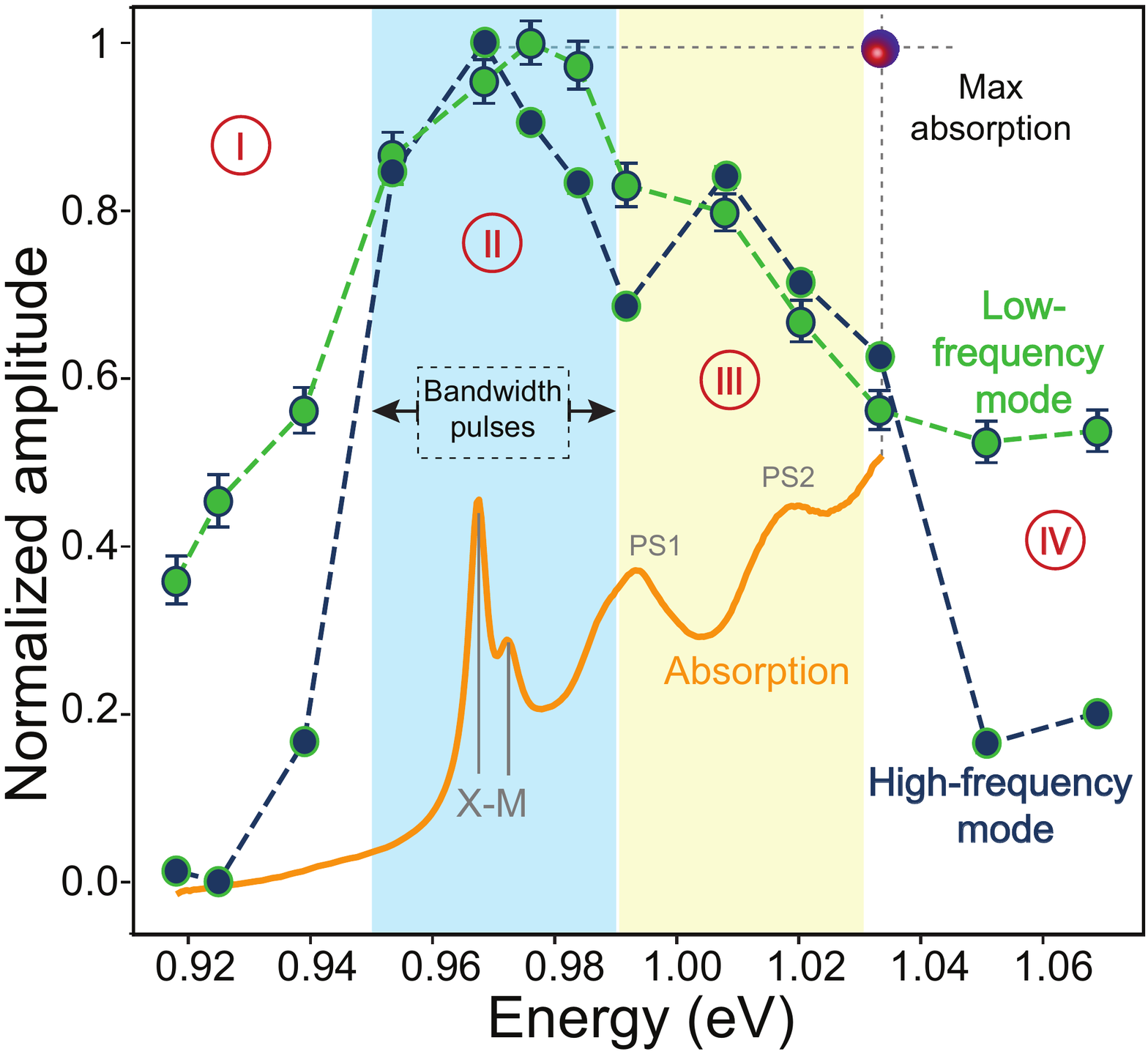}
	\caption{\footnotesize{Amplification and nonlinear coupling of the two magnon modes. The amplitudes are normalized on their maximum values. The experimentally accessed spectral range is divided into four regions, as described in details in the main text.}}
	\label{fig:Fig5}
	\end{figure}

Hence we analysed the spectral dependence of the amplitude of both magnon modes~\cite{SM}. The results of the data processing are shown in Fig. \ref{fig:Fig5}. We first discuss the trend of the hf-mode. The off-resonant photo-excitation (I) does not induce THz magnons, implying that an impulsive stimulated Raman generation of the hf-mode is not observed. The amplitude of the hf-mode increases steeply as the spectral range containing the X-M (II) is covered by the spectrum of the pump pulses. A comparison of the spectral dependence of the hf-mode with the absorption spectrum of NiO (both plotted in Fig. \ref{fig:Fig5}) reveals the amplification of the magnon mode to occur in a broader spectral range than the X-M itself. This behaviour is due to the bandwidth of the pump pulses, which being  ultrashort ($\approx$50 fs) are intrinsically broadband ($\approx$ 40 meV). As the pump photon-energy is further increased (III), so that the X-M is not directly induced anymore, the amplitude of the hf-mode is reduced. Two other transitions (PS1 and PS2) are photo-induced in spectral region III: they are phonon sidebands of the excitonic peak\cite{MironovaUlmane:2012ca}. Therefore a resonant pumping of such sidebands results again in inducing both the excitonic peak and the X-M,  as recently experimentally demonstrated\cite{Bossini:2018db}. However, the phonon sidebands implies a stronger optical absorption, so that the overall amplitude of the hf-mode is reduced in comparison with spectral range II, since in region III a portion of the pump photons are absorbed by the lattice. Finally, in region IV the X-M is not induced anymore and, accordingly, the amplitude of the hf-mode decreases to an almost vanishing value. We conclude that the pumping of the X-M unambiguously amplifies the hf-mode. The mechanism is a resonant drive but it is not purely dissipative, implying that pumping the material with photon energy corresponding to the maximum absorption (red dot in Fig. \ref{fig:Fig5})  does not deliver the most intense magnonic oscillations. This result is achieved if the X-M is resonantly driven.

We now turn the discussion to the lf-mode. Under non-resonant pumping (I and IV) the amplitude of the lf-mode is not negligible, consistently with the literature\cite{Tzschaschel:2017eh}. The observation of coherent oscillations in a multidomain state of the sample is ascribed to the different amplitudes of the lf-oscillations photo-induced in different T  domains. The ISRS excitation is conventionally described in terms of a light-induced effective field, which torques the spins triggering spin precession~\cite{Kimel2005,Kirilyuk2010}. The torque is maximum in T domains in which spins lie along a directional orthogonal to the effective field, implying that different T domains differently contribute to the overall detected magneto-optical response, which is hence not averaged out. Unexpectedly, also the lf-mode experiences an amplification in spectral regions II and III, although not as pronounced as the THz mode. The lf-mode is not reported to be involved in the X-M transition\cite{MironovaUlmane:2012ca}. We stress that the observed behaviour is strongly surprising and atypical. Experiments have demonstrated that the amplitude of magnonic oscillations induced by ISRS in a single-domain antiferromagnet, are not affected by a modification of the pump photon-energy in a spectral range lying below the band-gap~\cite{Bossini2014}.

A key characteristic of our NiO samples is that they are in a multidomain state. Since S domains can be neglected\cite{Tzschaschel:2017eh,SM} the T domains have to be taken into account. Hence, we formulate a macroscopic model. In particular, we consider two T domains (T1 and T2) and a wall separating them, and thus the unit vectors representing the two modes ($\mathbf{e}_\mathrm{hf}$ and $\mathbf{e}_\mathrm{lf}$) have different non-orthogonal orientations in the T1 and T2 domains. In the presence of such a domain wall the two modes can couple, due to the exchange interaction between the N\'eel vectors of the two different domains. In addition, since the wall breaks the translational symmetry, the eigenstates in one domain are not eigenstates in the other one. In a single T-domain and in the linear regime the two modes cannot couple, because they have different frequencies and are two different eigenstates of the Hamiltonian of the magnetic system.
As a starting point, we assume that the photo-excitation induces the hf-mode in a T1 ($\xi<0$, where $\xi$ is introduced in Fig. \ref{fig:Fig2}(a)) domain: $\mathbf{n}=a \mathbf{e}_\mathrm{hf}(\mathrm{T1}) \exp\left(i\omega_\mathrm{hf}t\right)$, where $a$ is the amplitude of the photo-induced magnons and depends on the intensity of the pump beam and on the cross-section of the X-M process. We model then how the hf-magnons in T1 can interact with the lf-magnon mode in T2 ($\xi\ge 0$), whose amplitude is represented in the following by $b$, via the domain wall. Considering the standard equations for the dynamics of $\mathbf{n}$~\cite{BarYakhtar:1980dz}, we obtain:

	\begin{eqnarray}\label{eq_magnetic_dynamics}
\mathbf{n}&\times&\left(\ddot{\mathbf{n}}-c^2\Delta \mathbf{n}+\gamma^2H_\mathrm{ex}\mathbf{H}_\mathrm{an}\right)\\
&=&\omega_\mathrm{hf}^2a \mathbf{n}\times\mathbf{e}_\mathrm{hf}(\mathrm{T1}) \exp\left(i\omega_\mathrm{hf}t\right)\delta(\xi),\nonumber
	\end{eqnarray}
	
\noindent where $c$ is the limiting propagation velocity of magnons, $\gamma$ is the gyromagnetic ratio, $H_\mathrm{ex}$ is the exchange field, $\mathbf{H}_\mathrm{an}$ is the effective anisotropy field given as usual by $\mathbf{H}_\mathrm{an}=-\partial w_\mathrm{an}/\partial \mathbf{n}$, $w_\mathrm{an}$ being the magnetic anisotropy energy~\cite{SM}.

Equation (\ref{eq_magnetic_dynamics}) can be solved in the stationary case in the absence of pump pulses, giving the spatial orientation of the N\'eel vector from the T1 to the T2 domain across the wall,  $\mathbf{n}_0(\xi)=\cos\varphi_0\mathbf{e}_0(\xi)+\sin\varphi_0\mathbf{e}_\mathrm{lf}(\xi)$, parametrized with the function $\varphi_0(\xi)$ and space-dependent vectors~\cite{SM}. The stationary solution is represented in Fig. \ref{fig:Fig2}(a), which displays how the orientation of the N\'eel vector changes in the domain wall. In particular, at the center of the wall ($\xi \ \approx \ 0$) the $\mathbf{n}$ vectors in the two domains are oriented parallel to each other. We calculate ~\cite{SM} the magnon spectra in a two-domain sample and find that they exhibit not only propagating modes, but also localised modes that correspond to oscillations of the domain wall. The frequencies of the latter modes depend on both the magneto-elastic coupling and on the exchange interaction between the N\'eel vectors in the two T domains. The frequency of one of the localised modes, $\omega_\mathrm{DW}$, can be approximated as $\omega_\mathrm{DW}\approx \omega_\mathrm{hf}+\omega_\mathrm{lf}$~\cite{SM}. 

The experimentally observed coupling between magnon modes with different frequencies is an intrinsically nonlinear process, as an input signal at a given frequency ($\omega_\mathrm{hf}$) is converted into an output with a different frequency ($\omega_\mathrm{lf}$). We thus expand $\mathbf{n}$ in series of powers of small deflections $\delta\mathbf{n}$, around the stationary solution $\mathbf{n}_0(\xi)$, so that the lowest-order non-linearity is achieved. We represent further $\delta\mathbf{n}$ as a sum of three eigenmodes: the localized domain wall mode, the lf- and the hf-mode. Relying on the initial conditions, in which the domain wall is not in motion prior to the photo-excitation, the equations for the corresponding time-dependent amplitudes $b_\mathrm{DW}$, $b_\mathrm{lf}$, and $b_\mathrm{hf}$ in the T2 domain are~\cite{SM}

\begin{eqnarray}\label{eq_time_dependent}
	&&\ddot{b}_\mathrm{DW}+\frac{1}{\tau}\dot{b}_\mathrm{DW}+\omega_\mathrm{DW}^2b_\mathrm{DW}= \omega_\mathrm{hf}^2a\exp\left(i\omega_\mathrm{hf}t\right),\label{eq_time_dependenta}\nonumber\\
&&\ddot{b}_\mathrm{hf}+\frac{1}{\tau} \dot{b}_\mathrm{hf}+\omega_\mathrm{hf}^2b_\mathrm{hf}  = \frac{1}{3}\omega_\mathrm{hf}^2a\exp\left(i\omega_\mathrm{hf}t\right),\label{eq_time_dependentb}\\
&&\ddot{b}_\mathrm{lf}+\frac{1}{\tau} \dot{b}_\mathrm{lf}+\omega_\mathrm{lf}^2\left(1-4P{b}_\mathrm{DW}\right)b_\mathrm{lf} = \frac{1}{3}\omega_\mathrm{hf}^2a\exp\left(i\omega_\mathrm{hf}t\right),\nonumber \label{eq_time_dependentc}
\end{eqnarray}

\noindent where $1/\tau\ll\omega_\mathrm{lf},\omega_\mathrm{hf},\omega_\mathrm{DW}$ is the relaxation time due to the Gilbert damping and the coefficient $P$ describes the coupling between the lf- and domain wall-mode. Let us stress that the source terms in Eqs. (\ref{eq_time_dependent}) are photo-driven hf-magnons in the T1 domain. The relevant quantity to compute, for the sake of comparison with the experiment is thus $b_\mathrm{lf}$:

\begin{eqnarray}\label{eq_lf_mode}
b_\mathrm{lf}&=&\frac{\omega^2_\mathrm{hf}a}{\omega_\mathrm{lf}^2-\omega_\mathrm{hf}^2}\exp\left(i\omega_\mathrm{hf}t\right)\\
&+&\frac{4a^2\omega_\mathrm{lf}^2\omega^2_\mathrm{hf}e^{\lambda t}}{3(\omega_\mathrm{DW}^2-\omega_\mathrm{hf}^2)^2(\omega_\mathrm{lf}^2-\omega_\mathrm{hf}^2)} \exp[i\overbrace{\left(\omega_\mathrm{hf}-\omega_\mathrm{DW}\right)}^{- \omega_\mathrm{lf}}t],\nonumber
	\end{eqnarray}

\noindent where $\lambda=\sqrt{2}aP\omega^2_\mathrm{hf}/3-1/{\tau}$ is the instability parameter for parametric downconversion. The first term corresponds to a non-resonant excitations of lf-magnons with wave-vector high-enough, that they match the frequency of the hf-mode (i.e. 1.07 THz)~\cite{Bossini:2017bo}. These magnons may only contribute to the background signal, once they relax via scattering events. The second contribution originates from the nonlinear term in Eq. (\ref{eq_time_dependent}c) and corresponds to the beating between frequencies $\omega_\mathrm{DW}$ and $\omega_\mathrm{hf}$. Crucially, we note that quantitatively $\omega_\mathrm{hf}-\omega_\mathrm{DW}$ matches $\omega_\mathrm{lf}$. Moreover, since $\omega_\mathrm{lf}\approx 0.1 \ \omega_\mathrm{hf}\ll\omega_\mathrm{hf}$, we can reformulate the second contribution as

	\begin{equation}\label{eq_lf_mode_2}
b_\mathrm{lf}=\frac{a^2e^{\lambda t}}{3}\exp\left(i\omega_\mathrm{lf}t\right).
	\end{equation}
	
\noindent In essence our model describes an additional contribution to the lf-mode in the T2 domain, originating from the non-linear magneto-elastic interaction between the domain wall mode and the photo-induced hf-magnons in the T1 domain, activated by the hf-magnons hitting on the wall. This concept is consistent with the experimental observations reported in Fig. \ref{fig:Fig3}. The interpretation of our data in terms of the mechanism described in the model above is further substantiated by measurements performed on a NiO sample in a single-domain state, displaying amplification of the hf-mode only~\cite{SM}. Further details of the model have been recently published.~\cite{Gomonay2021}

We rule out the possibility that the lattice mediate the coupling between the modes, as no phonon with the required frequency for the frequency mixing are reported in the dispersion of NiO.~\cite{Reichardt:1975bm}

Our results suggest that a proper sample engineering may allow both an even more pronounced amplitude amplification of the oscillations of $\mathbf{n}$ and a coupling even among propagating magnons. The former phenomenon is relevant for achieving non-linearities in the spin dynamics, while the latter is a milestone towards the establishment of coherence- and energy-transfer between magnonic branches on the characteristic femtosecond time- and (sub)-micrometer length-scales of collective spin eigenmodes. \nocite{Stancik2008,Sell1968,Sell:1967jo,Macfarlane1971,Polley:2018cm,Satoh2010Japp,Herrmann_Ronzaud_1978,Poeschel1933,Cooper1995}

{\footnotesize{This work was supported by the Deutsche Forschungsgemeinschaft through the International Collaborative Research Centre TRR160 (Project B9),  the DFG programme BO 5074/1-1 and UH 90/13-1, by the COST Action MAGNETOFON (grant number CA17123, STSM number CA17123-44749).  M.B. and L.S. acknowledge support from the Knut and Alice Wallenberg Foundation, grant 2017.0158. SB acknowledges support from the Swedish Research Council (VR), Grant: 2018-04611. O.G. acknowledges funding by the Deutsche Forschungsgemeinschaft (DFG, German Research Foundation) - TRR 173 - 268565370 (project A11), TRR 288 - 422213477 (project A09), and project SHARP 397322108, the ERC Synergy Grant SC2 (No. 610115), and partial support by the National Science Foundation under Grant No. NSF PHY-1748958. She also acknowledges discussions with Oleg Tchernyshyov. T.S. ackowledges H. Ueda for providing the single-domain sample.}}


\begin{thebibliography}{45}%
\makeatletter
\providecommand \@ifxundefined [1]{%
 \@ifx{#1\undefined}
}%
\providecommand \@ifnum [1]{%
 \ifnum #1\expandafter \@firstoftwo
 \else \expandafter \@secondoftwo
 \fi
}%
\providecommand \@ifx [1]{%
 \ifx #1\expandafter \@firstoftwo
 \else \expandafter \@secondoftwo
 \fi
}%
\providecommand \natexlab [1]{#1}%
\providecommand \enquote  [1]{``#1''}%
\providecommand \bibnamefont  [1]{#1}%
\providecommand \bibfnamefont [1]{#1}%
\providecommand \citenamefont [1]{#1}%
\providecommand \href@noop [0]{\@secondoftwo}%
\providecommand \href [0]{\begingroup \@sanitize@url \@href}%
\providecommand \@href[1]{\@@startlink{#1}\@@href}%
\providecommand \@@href[1]{\endgroup#1\@@endlink}%
\providecommand \@sanitize@url [0]{\catcode `\\12\catcode `\$12\catcode
  `\&12\catcode `\#12\catcode `\^12\catcode `\_12\catcode `\%12\relax}%
\providecommand \@@startlink[1]{}%
\providecommand \@@endlink[0]{}%
\providecommand \url  [0]{\begingroup\@sanitize@url \@url }%
\providecommand \@url [1]{\endgroup\@href {#1}{\urlprefix }}%
\providecommand \urlprefix  [0]{URL }%
\providecommand \Eprint [0]{\href }%
\providecommand \doibase [0]{http://dx.doi.org/}%
\providecommand \selectlanguage [0]{\@gobble}%
\providecommand \bibinfo  [0]{\@secondoftwo}%
\providecommand \bibfield  [0]{\@secondoftwo}%
\providecommand \translation [1]{[#1]}%
\providecommand \BibitemOpen [0]{}%
\providecommand \bibitemStop [0]{}%
\providecommand \bibitemNoStop [0]{.\EOS\space}%
\providecommand \EOS [0]{\spacefactor3000\relax}%
\providecommand \BibitemShut  [1]{\csname bibitem#1\endcsname}%
\let\auto@bib@innerbib\@empty
\bibitem [{\citenamefont {Wadley}\ \emph {et~al.}(2016)\citenamefont {Wadley},
  \citenamefont {Howells}, \citenamefont {{\v Z}elezn{\'{y}}}, \citenamefont
  {Andrews}, \citenamefont {Hills}, \citenamefont {Campion}, \citenamefont
  {Nov\'{a}k}, \citenamefont {Olejn\'{i}k}, \citenamefont {Maccherozzi},
  \citenamefont {Dhesi}, \citenamefont {Martin}, \citenamefont {Wagner},
  \citenamefont {Wunderlich}, \citenamefont {Freimuth}, \citenamefont
  {Mokrousov}, \citenamefont {Kune{\v s}}, \citenamefont {Chauhan},
  \citenamefont {Grzybowski}, \citenamefont {Rushforth}, \citenamefont
  {Edmonds}, \citenamefont {Gallagher},\ and\ \citenamefont
  {Jungwirth}}]{Wadley2016a}%
  \BibitemOpen
  \bibfield  {author} {\bibinfo {author} {\bibfnamefont {P.}~\bibnamefont
  {Wadley}}, \bibinfo {author} {\bibfnamefont {B.}~\bibnamefont {Howells}},
  \bibinfo {author} {\bibfnamefont {J.}~\bibnamefont {{\v Z}elezn{\'{y}}}},
  \bibinfo {author} {\bibfnamefont {C.}~\bibnamefont {Andrews}}, \bibinfo
  {author} {\bibfnamefont {V.}~\bibnamefont {Hills}}, \bibinfo {author}
  {\bibfnamefont {R.~P.}\ \bibnamefont {Campion}}, \bibinfo {author}
  {\bibfnamefont {V.}~\bibnamefont {Nov\'{a}k}}, \bibinfo {author}
  {\bibfnamefont {K.}~\bibnamefont {Olejn\'{i}k}}, \bibinfo {author}
  {\bibfnamefont {F.}~\bibnamefont {Maccherozzi}}, \bibinfo {author}
  {\bibfnamefont {S.~S.}\ \bibnamefont {Dhesi}}, \bibinfo {author}
  {\bibfnamefont {S.~Y.}\ \bibnamefont {Martin}}, \bibinfo {author}
  {\bibfnamefont {T.}~\bibnamefont {Wagner}}, \bibinfo {author} {\bibfnamefont
  {J.}~\bibnamefont {Wunderlich}}, \bibinfo {author} {\bibfnamefont
  {F.}~\bibnamefont {Freimuth}}, \bibinfo {author} {\bibfnamefont
  {Y.}~\bibnamefont {Mokrousov}}, \bibinfo {author} {\bibfnamefont
  {J.}~\bibnamefont {Kune{\v s}}}, \bibinfo {author} {\bibfnamefont {J.~S.}\
  \bibnamefont {Chauhan}}, \bibinfo {author} {\bibfnamefont {M.~J.}\
  \bibnamefont {Grzybowski}}, \bibinfo {author} {\bibfnamefont {A.~W.}\
  \bibnamefont {Rushforth}}, \bibinfo {author} {\bibfnamefont {K.~W.}\
  \bibnamefont {Edmonds}}, \bibinfo {author} {\bibfnamefont {B.~L.}\
  \bibnamefont {Gallagher}}, \ and\ \bibinfo {author} {\bibfnamefont
  {T.}~\bibnamefont {Jungwirth}},\ }\href {\doibase 10.1126/science.aab1031}
  {\bibfield  {journal} {\bibinfo  {journal} {Science}\ }\textbf {\bibinfo
  {volume} {351}},\ \bibinfo {pages} {587} (\bibinfo {year}
  {2016})}\BibitemShut {NoStop}%
\bibitem [{\citenamefont {Lebrun}\ \emph {et~al.}(2018)\citenamefont {Lebrun},
  \citenamefont {Ross}, \citenamefont {Bender}, \citenamefont {Qaiumzadeh},
  \citenamefont {Baldrati}, \citenamefont {Cramer}, \citenamefont {Brataas},
  \citenamefont {Duine},\ and\ \citenamefont {Kl{\"a}ui}}]{Lebrun:2018kt}%
  \BibitemOpen
  \bibfield  {author} {\bibinfo {author} {\bibfnamefont {R.}~\bibnamefont
  {Lebrun}}, \bibinfo {author} {\bibfnamefont {A.}~\bibnamefont {Ross}},
  \bibinfo {author} {\bibfnamefont {S.~A.}\ \bibnamefont {Bender}}, \bibinfo
  {author} {\bibfnamefont {A.}~\bibnamefont {Qaiumzadeh}}, \bibinfo {author}
  {\bibfnamefont {L.}~\bibnamefont {Baldrati}}, \bibinfo {author}
  {\bibfnamefont {J.}~\bibnamefont {Cramer}}, \bibinfo {author} {\bibfnamefont
  {A.}~\bibnamefont {Brataas}}, \bibinfo {author} {\bibfnamefont {R.~A.}\
  \bibnamefont {Duine}}, \ and\ \bibinfo {author} {\bibfnamefont
  {M.}~\bibnamefont {Kl{\"a}ui}},\ }\href {\doibase 10.1038/s41586-018-0490-7}
  {\bibfield  {journal} {\bibinfo  {journal} {Nature}\ }\textbf {\bibinfo
  {volume} {561}},\ \bibinfo {pages} {222} (\bibinfo {year}
  {2018})}\BibitemShut {NoStop}%
\bibitem [{\citenamefont {Baltz}\ \emph {et~al.}(2018)\citenamefont {Baltz},
  \citenamefont {Manchon}, \citenamefont {Tsoi}, \citenamefont {Moriyama},
  \citenamefont {Ono},\ and\ \citenamefont {Tserkovnyak}}]{Baltz:2018iv}%
  \BibitemOpen
  \bibfield  {author} {\bibinfo {author} {\bibfnamefont {V.}~\bibnamefont
  {Baltz}}, \bibinfo {author} {\bibfnamefont {A.}~\bibnamefont {Manchon}},
  \bibinfo {author} {\bibfnamefont {M.}~\bibnamefont {Tsoi}}, \bibinfo {author}
  {\bibfnamefont {T.}~\bibnamefont {Moriyama}}, \bibinfo {author}
  {\bibfnamefont {T.}~\bibnamefont {Ono}}, \ and\ \bibinfo {author}
  {\bibfnamefont {Y.}~\bibnamefont {Tserkovnyak}},\ }\href {\doibase
  10.1103/RevModPhys.90.015005} {\bibfield  {journal} {\bibinfo  {journal}
  {Reviews of Modern Physics}\ }\textbf {\bibinfo {volume} {90}},\ \bibinfo
  {pages} {015005} (\bibinfo {year} {2018})}\BibitemShut {NoStop}%
\bibitem [{\citenamefont {Gomonay}\ \emph {et~al.}(2018)\citenamefont
  {Gomonay}, \citenamefont {Baltz}, \citenamefont {Brataas},\ and\
  \citenamefont {Tserkovnyak}}]{Gomonay:2018js}%
  \BibitemOpen
  \bibfield  {author} {\bibinfo {author} {\bibfnamefont {O.}~\bibnamefont
  {Gomonay}}, \bibinfo {author} {\bibfnamefont {V.}~\bibnamefont {Baltz}},
  \bibinfo {author} {\bibfnamefont {A.}~\bibnamefont {Brataas}}, \ and\
  \bibinfo {author} {\bibfnamefont {Y.}~\bibnamefont {Tserkovnyak}},\ }\href
  {\doibase 10.1038/s41567-018-0049-4} {\bibfield  {journal} {\bibinfo
  {journal} {Nature Physics}\ }\textbf {\bibinfo {volume} {14}},\ \bibinfo
  {pages} {213} (\bibinfo {year} {2018})}\BibitemShut {NoStop}%
\bibitem [{\citenamefont {Baldrati}\ \emph {et~al.}(2020)\citenamefont
  {Baldrati}, \citenamefont {Schmitt}, \citenamefont {Gomonay}, \citenamefont
  {Lebrun}, \citenamefont {Ramos}, \citenamefont {Saitoh}, \citenamefont
  {Sinova},\ and\ \citenamefont {Kl{\"a}ui}}]{Baldrati2020}%
  \BibitemOpen
  \bibfield  {author} {\bibinfo {author} {\bibfnamefont {L.}~\bibnamefont
  {Baldrati}}, \bibinfo {author} {\bibfnamefont {C.}~\bibnamefont {Schmitt}},
  \bibinfo {author} {\bibfnamefont {O.}~\bibnamefont {Gomonay}}, \bibinfo
  {author} {\bibfnamefont {R.}~\bibnamefont {Lebrun}}, \bibinfo {author}
  {\bibfnamefont {R.}~\bibnamefont {Ramos}}, \bibinfo {author} {\bibfnamefont
  {E.}~\bibnamefont {Saitoh}}, \bibinfo {author} {\bibfnamefont
  {J.}~\bibnamefont {Sinova}}, \ and\ \bibinfo {author} {\bibfnamefont
  {M.}~\bibnamefont {Kl{\"a}ui}},\ }\href {\doibase
  10.1103/PhysRevLett.125.077201} {\bibfield  {journal} {\bibinfo  {journal}
  {Phys. Rev. Lett.}\ }\textbf {\bibinfo {volume} {125}},\ \bibinfo {pages}
  {077201} (\bibinfo {year} {2020})}\BibitemShut {NoStop}%
\bibitem [{\citenamefont {Meer}\ \emph {et~al.}(2020)\citenamefont {Meer},
  \citenamefont {Schreiber}, \citenamefont {Schmitt}, \citenamefont {Ramos},
  \citenamefont {Saitoh}, \citenamefont {Gomonay}, \citenamefont {Sinova},
  \citenamefont {Baldrati},\ and\ \citenamefont {Kl\"{a}ui}}]{Meer2020}%
  \BibitemOpen
  \bibfield  {author} {\bibinfo {author} {\bibfnamefont {H.}~\bibnamefont
  {Meer}}, \bibinfo {author} {\bibfnamefont {F.}~\bibnamefont {Schreiber}},
  \bibinfo {author} {\bibfnamefont {C.}~\bibnamefont {Schmitt}}, \bibinfo
  {author} {\bibfnamefont {R.}~\bibnamefont {Ramos}}, \bibinfo {author}
  {\bibfnamefont {E.}~\bibnamefont {Saitoh}}, \bibinfo {author} {\bibfnamefont
  {O.}~\bibnamefont {Gomonay}}, \bibinfo {author} {\bibfnamefont
  {J.}~\bibnamefont {Sinova}}, \bibinfo {author} {\bibfnamefont
  {L.}~\bibnamefont {Baldrati}}, \ and\ \bibinfo {author} {\bibfnamefont
  {M.}~\bibnamefont {Kl\"{a}ui}},\ }\href {\doibase
  10.1021/acs.nanolett.0c03367} {\bibfield  {journal} {\bibinfo  {journal}
  {Nano Letters}\ }\textbf {\bibinfo {volume} {21}},\ \bibinfo {pages} {114}
  (\bibinfo {year} {2020})}\BibitemShut {NoStop}%
\bibitem [{\citenamefont {Zhang}\ \emph {et~al.}(2019)\citenamefont {Zhang},
  \citenamefont {Finley}, \citenamefont {Safi},\ and\ \citenamefont
  {Liu}}]{Zhang:2019gc}%
  \BibitemOpen
  \bibfield  {author} {\bibinfo {author} {\bibfnamefont {P.}~\bibnamefont
  {Zhang}}, \bibinfo {author} {\bibfnamefont {J.}~\bibnamefont {Finley}},
  \bibinfo {author} {\bibfnamefont {T.}~\bibnamefont {Safi}}, \ and\ \bibinfo
  {author} {\bibfnamefont {L.}~\bibnamefont {Liu}},\ }\href {\doibase
  10.1103/PhysRevLett.123.247206} {\bibfield  {journal} {\bibinfo  {journal}
  {Phys. Rev. Lett.}\ }\textbf {\bibinfo {volume} {123}},\ \bibinfo {pages}
  {247206} (\bibinfo {year} {2019})}\BibitemShut {NoStop}%
\bibitem [{\citenamefont {N\v{e}mec}\ \emph {et~al.}(2018)\citenamefont
  {N\v{e}mec}, \citenamefont {Fiebig}, \citenamefont {Kampfrath},\ and\
  \citenamefont {Kimel}}]{Nemec:2018gw}%
  \BibitemOpen
  \bibfield  {author} {\bibinfo {author} {\bibfnamefont {P.}~\bibnamefont
  {N\v{e}mec}}, \bibinfo {author} {\bibfnamefont {M.}~\bibnamefont {Fiebig}},
  \bibinfo {author} {\bibfnamefont {T.}~\bibnamefont {Kampfrath}}, \ and\
  \bibinfo {author} {\bibfnamefont {A.~V.}\ \bibnamefont {Kimel}},\ }\href
  {\doibase 10.1038/s41567-018-0051-x} {\bibfield  {journal} {\bibinfo
  {journal} {Nature Physics}\ }\textbf {\bibinfo {volume} {14}},\ \bibinfo
  {pages} {229} (\bibinfo {year} {2018})}\BibitemShut {NoStop}%
\bibitem [{\citenamefont {Satoh}\ \emph
  {et~al.}(2010{\natexlab{a}})\citenamefont {Satoh}, \citenamefont {Cho},
  \citenamefont {Iida}, \citenamefont {Shimura}, \citenamefont {Kuroda},
  \citenamefont {Ueda}, \citenamefont {Ueda}, \citenamefont {Ivanov},
  \citenamefont {Nori},\ and\ \citenamefont {Fiebig}}]{Satoh2010}%
  \BibitemOpen
  \bibfield  {author} {\bibinfo {author} {\bibfnamefont {T.}~\bibnamefont
  {Satoh}}, \bibinfo {author} {\bibfnamefont {S.-J.}\ \bibnamefont {Cho}},
  \bibinfo {author} {\bibfnamefont {R.}~\bibnamefont {Iida}}, \bibinfo {author}
  {\bibfnamefont {T.}~\bibnamefont {Shimura}}, \bibinfo {author} {\bibfnamefont
  {K.}~\bibnamefont {Kuroda}}, \bibinfo {author} {\bibfnamefont
  {H.}~\bibnamefont {Ueda}}, \bibinfo {author} {\bibfnamefont {Y.}~\bibnamefont
  {Ueda}}, \bibinfo {author} {\bibfnamefont {B.~A.}\ \bibnamefont {Ivanov}},
  \bibinfo {author} {\bibfnamefont {F.}~\bibnamefont {Nori}}, \ and\ \bibinfo
  {author} {\bibfnamefont {M.}~\bibnamefont {Fiebig}},\ }\href {\doibase
  10.1103/PhysRevLett.105.077402} {\bibfield  {journal} {\bibinfo  {journal}
  {Phys. Rev. Lett.}\ }\textbf {\bibinfo {volume} {105}},\ \bibinfo {pages}
  {077402} (\bibinfo {year} {2010}{\natexlab{a}})}\BibitemShut {NoStop}%
\bibitem [{\citenamefont {Baierl}\ \emph {et~al.}(2016)\citenamefont {Baierl},
  \citenamefont {Mentink}, \citenamefont {Hohenleutner}, \citenamefont {Braun},
  \citenamefont {Do}, \citenamefont {Lange}, \citenamefont {Sell},
  \citenamefont {Fiebig}, \citenamefont {Woltersdorf}, \citenamefont
  {Kampfrath},\ and\ \citenamefont {Huber}}]{Baierl:2016cb}%
  \BibitemOpen
  \bibfield  {author} {\bibinfo {author} {\bibfnamefont {S.}~\bibnamefont
  {Baierl}}, \bibinfo {author} {\bibfnamefont {J.~H.}\ \bibnamefont {Mentink}},
  \bibinfo {author} {\bibfnamefont {M.}~\bibnamefont {Hohenleutner}}, \bibinfo
  {author} {\bibfnamefont {L.}~\bibnamefont {Braun}}, \bibinfo {author}
  {\bibfnamefont {T.-M.}\ \bibnamefont {Do}}, \bibinfo {author} {\bibfnamefont
  {C.}~\bibnamefont {Lange}}, \bibinfo {author} {\bibfnamefont
  {A.}~\bibnamefont {Sell}}, \bibinfo {author} {\bibfnamefont {M.}~\bibnamefont
  {Fiebig}}, \bibinfo {author} {\bibfnamefont {G.}~\bibnamefont {Woltersdorf}},
  \bibinfo {author} {\bibfnamefont {T.}~\bibnamefont {Kampfrath}}, \ and\
  \bibinfo {author} {\bibfnamefont {R.}~\bibnamefont {Huber}},\ }\href
  {\doibase 10.1103/physrevlett.117.197201} {\bibfield  {journal} {\bibinfo
  {journal} {Phys. Rev. Lett.}\ }\textbf {\bibinfo {volume} {117}},\ \bibinfo
  {pages} {197201} (\bibinfo {year} {2016})}\BibitemShut {NoStop}%
\bibitem [{\citenamefont {Kampfrath}\ \emph {et~al.}(2010)\citenamefont
  {Kampfrath}, \citenamefont {Sell}, \citenamefont {Klatt}, \citenamefont
  {Pashkin}, \citenamefont {M{\"a}hrlein}, \citenamefont {Dekorsy},
  \citenamefont {Wolf}, \citenamefont {Fiebig}, \citenamefont {Leitenstorfer},\
  and\ \citenamefont {Huber}}]{Kampfrath2010}%
  \BibitemOpen
  \bibfield  {author} {\bibinfo {author} {\bibfnamefont {T.}~\bibnamefont
  {Kampfrath}}, \bibinfo {author} {\bibfnamefont {A.}~\bibnamefont {Sell}},
  \bibinfo {author} {\bibfnamefont {G.}~\bibnamefont {Klatt}}, \bibinfo
  {author} {\bibfnamefont {A.}~\bibnamefont {Pashkin}}, \bibinfo {author}
  {\bibfnamefont {S.}~\bibnamefont {M{\"a}hrlein}}, \bibinfo {author}
  {\bibfnamefont {T.}~\bibnamefont {Dekorsy}}, \bibinfo {author} {\bibfnamefont
  {M.}~\bibnamefont {Wolf}}, \bibinfo {author} {\bibfnamefont {M.}~\bibnamefont
  {Fiebig}}, \bibinfo {author} {\bibfnamefont {A.}~\bibnamefont
  {Leitenstorfer}}, \ and\ \bibinfo {author} {\bibfnamefont {R.}~\bibnamefont
  {Huber}},\ }\href {\doibase 10.1038/nphoton.2010.259} {\bibfield  {journal}
  {\bibinfo  {journal} {Nature Photonics}\ }\textbf {\bibinfo {volume} {5}},\
  \bibinfo {pages} {31} (\bibinfo {year} {2010})}\BibitemShut {NoStop}%
\bibitem [{\citenamefont {Bossini}\ \emph {et~al.}(2016)\citenamefont
  {Bossini}, \citenamefont {Dal~Conte}, \citenamefont {Hashimoto},
  \citenamefont {Secchi}, \citenamefont {Pisarev}, \citenamefont {Rasing},
  \citenamefont {Cerullo},\ and\ \citenamefont {Kimel}}]{Bossini2016}%
  \BibitemOpen
  \bibfield  {author} {\bibinfo {author} {\bibfnamefont {D.}~\bibnamefont
  {Bossini}}, \bibinfo {author} {\bibfnamefont {S.}~\bibnamefont {Dal~Conte}},
  \bibinfo {author} {\bibfnamefont {Y.}~\bibnamefont {Hashimoto}}, \bibinfo
  {author} {\bibfnamefont {A.}~\bibnamefont {Secchi}}, \bibinfo {author}
  {\bibfnamefont {R.~V.}\ \bibnamefont {Pisarev}}, \bibinfo {author}
  {\bibfnamefont {{\relax Th}.}~\bibnamefont {Rasing}}, \bibinfo {author}
  {\bibfnamefont {G.}~\bibnamefont {Cerullo}}, \ and\ \bibinfo {author}
  {\bibfnamefont {A.~V.}\ \bibnamefont {Kimel}},\ }\href {\doibase
  10.1038/ncomms10645} {\bibfield  {journal} {\bibinfo  {journal} {Nature
  Communications}\ }\textbf {\bibinfo {volume} {7}},\ \bibinfo {pages} {10645}
  (\bibinfo {year} {2016})}\BibitemShut {NoStop}%
\bibitem [{\citenamefont {Tzschaschel}\ \emph {et~al.}(2017)\citenamefont
  {Tzschaschel}, \citenamefont {Otani}, \citenamefont {Iida}, \citenamefont
  {Shimura}, \citenamefont {Ueda}, \citenamefont {G{\"u}nther}, \citenamefont
  {Fiebig},\ and\ \citenamefont {Satoh}}]{Tzschaschel:2017eh}%
  \BibitemOpen
  \bibfield  {author} {\bibinfo {author} {\bibfnamefont {C.}~\bibnamefont
  {Tzschaschel}}, \bibinfo {author} {\bibfnamefont {K.}~\bibnamefont {Otani}},
  \bibinfo {author} {\bibfnamefont {R.}~\bibnamefont {Iida}}, \bibinfo {author}
  {\bibfnamefont {T.}~\bibnamefont {Shimura}}, \bibinfo {author} {\bibfnamefont
  {H.}~\bibnamefont {Ueda}}, \bibinfo {author} {\bibfnamefont {S.}~\bibnamefont
  {G{\"u}nther}}, \bibinfo {author} {\bibfnamefont {M.}~\bibnamefont {Fiebig}},
  \ and\ \bibinfo {author} {\bibfnamefont {T.}~\bibnamefont {Satoh}},\ }\href
  {\doibase 10.1103/PhysRevB.95.174407} {\bibfield  {journal} {\bibinfo
  {journal} {Physical Review B}\ }\textbf {\bibinfo {volume} {95}},\ \bibinfo
  {pages} {174407} (\bibinfo {year} {2017})}\BibitemShut {NoStop}%
\bibitem [{\citenamefont {Nishitani}\ \emph {et~al.}(2010)\citenamefont
  {Nishitani}, \citenamefont {Kozuki}, \citenamefont {Nagashima},\ and\
  \citenamefont {Hangyo}}]{Nishitani2010}%
  \BibitemOpen
  \bibfield  {author} {\bibinfo {author} {\bibfnamefont {J.}~\bibnamefont
  {Nishitani}}, \bibinfo {author} {\bibfnamefont {K.}~\bibnamefont {Kozuki}},
  \bibinfo {author} {\bibfnamefont {T.}~\bibnamefont {Nagashima}}, \ and\
  \bibinfo {author} {\bibfnamefont {M.}~\bibnamefont {Hangyo}},\ }\href
  {\doibase 10.1063/1.3436635} {\bibfield  {journal} {\bibinfo  {journal}
  {Applied Physics Letters}\ }\textbf {\bibinfo {volume} {96}},\ \bibinfo
  {pages} {221906} (\bibinfo {year} {2010})}\BibitemShut {NoStop}%
\bibitem [{\citenamefont {Simoncig}\ \emph {et~al.}(2017)\citenamefont
  {Simoncig}, \citenamefont {Mincigrucci}, \citenamefont {Principi},
  \citenamefont {Bencivenga}, \citenamefont {Calvi}, \citenamefont {Foglia},
  \citenamefont {Kurdi}, \citenamefont {Matruglio}, \citenamefont {Zilio},
  \citenamefont {Masciotti}, \citenamefont {Lazzarino},\ and\ \citenamefont
  {Masciovecchio}}]{Simoncig:2017bn}%
  \BibitemOpen
  \bibfield  {author} {\bibinfo {author} {\bibfnamefont {A.}~\bibnamefont
  {Simoncig}}, \bibinfo {author} {\bibfnamefont {R.}~\bibnamefont
  {Mincigrucci}}, \bibinfo {author} {\bibfnamefont {E.}~\bibnamefont
  {Principi}}, \bibinfo {author} {\bibfnamefont {F.}~\bibnamefont
  {Bencivenga}}, \bibinfo {author} {\bibfnamefont {A.}~\bibnamefont {Calvi}},
  \bibinfo {author} {\bibfnamefont {L.}~\bibnamefont {Foglia}}, \bibinfo
  {author} {\bibfnamefont {G.}~\bibnamefont {Kurdi}}, \bibinfo {author}
  {\bibfnamefont {A.}~\bibnamefont {Matruglio}}, \bibinfo {author}
  {\bibfnamefont {S.~{\relax Dal}.}\ \bibnamefont {Zilio}}, \bibinfo {author}
  {\bibfnamefont {V.}~\bibnamefont {Masciotti}}, \bibinfo {author}
  {\bibfnamefont {M.}~\bibnamefont {Lazzarino}}, \ and\ \bibinfo {author}
  {\bibfnamefont {C.}~\bibnamefont {Masciovecchio}},\ }\href {\doibase
  10.1103/PhysRevMaterials.1.073802} {\bibfield  {journal} {\bibinfo  {journal}
  {Physical Review Materials}\ }\textbf {\bibinfo {volume} {1}},\ \bibinfo
  {pages} {073802} (\bibinfo {year} {2017})}\BibitemShut {NoStop}%
\bibitem [{\citenamefont {Kanda}\ \emph {et~al.}(2011)\citenamefont {Kanda},
  \citenamefont {Higuchi}, \citenamefont {Shimizu}, \citenamefont {Konishi},
  \citenamefont {Yoshioka},\ and\ \citenamefont
  {Kuwata-Gonokami}}]{Kanda:2011ja}%
  \BibitemOpen
  \bibfield  {author} {\bibinfo {author} {\bibfnamefont {N.}~\bibnamefont
  {Kanda}}, \bibinfo {author} {\bibfnamefont {T.}~\bibnamefont {Higuchi}},
  \bibinfo {author} {\bibfnamefont {H.}~\bibnamefont {Shimizu}}, \bibinfo
  {author} {\bibfnamefont {K.}~\bibnamefont {Konishi}}, \bibinfo {author}
  {\bibfnamefont {K.}~\bibnamefont {Yoshioka}}, \ and\ \bibinfo {author}
  {\bibfnamefont {M.}~\bibnamefont {Kuwata-Gonokami}},\ }\href {\doibase
  10.1038/ncomms1366} {\bibfield  {journal} {\bibinfo  {journal} {Nature
  Communications}\ }\textbf {\bibinfo {volume} {2}},\ \bibinfo {pages} {362}
  (\bibinfo {year} {2011})}\BibitemShut {NoStop}%
\bibitem [{\citenamefont {Bossini}\ \emph {et~al.}(2019)\citenamefont
  {Bossini}, \citenamefont {Conte}, \citenamefont {Cerullo}, \citenamefont
  {Gomonay}, \citenamefont {Pisarev}, \citenamefont {Borovsak}, \citenamefont
  {Mihailovic}, \citenamefont {Sinova}, \citenamefont {Mentink}, \citenamefont
  {Rasing},\ and\ \citenamefont {Kimel}}]{Bossini:2019in}%
  \BibitemOpen
  \bibfield  {author} {\bibinfo {author} {\bibfnamefont {D.}~\bibnamefont
  {Bossini}}, \bibinfo {author} {\bibfnamefont {S.~{\relax Dal}.}\ \bibnamefont
  {Conte}}, \bibinfo {author} {\bibfnamefont {G.}~\bibnamefont {Cerullo}},
  \bibinfo {author} {\bibfnamefont {O.}~\bibnamefont {Gomonay}}, \bibinfo
  {author} {\bibfnamefont {R.~V.}\ \bibnamefont {Pisarev}}, \bibinfo {author}
  {\bibfnamefont {M.}~\bibnamefont {Borovsak}}, \bibinfo {author}
  {\bibfnamefont {D.}~\bibnamefont {Mihailovic}}, \bibinfo {author}
  {\bibfnamefont {J.}~\bibnamefont {Sinova}}, \bibinfo {author} {\bibfnamefont
  {J.~H.}\ \bibnamefont {Mentink}}, \bibinfo {author} {\bibfnamefont {{\relax
  Th}.}~\bibnamefont {Rasing}}, \ and\ \bibinfo {author} {\bibfnamefont
  {A.~V.}\ \bibnamefont {Kimel}},\ }\href {\doibase
  10.1103/PhysRevB.100.024428} {\bibfield  {journal} {\bibinfo  {journal}
  {Physical Review B}\ }\textbf {\bibinfo {volume} {100}},\ \bibinfo {pages}
  {024428} (\bibinfo {year} {2019})}\BibitemShut {NoStop}%
\bibitem [{\citenamefont {Baruchel}\ \emph {et~al.}(1977)\citenamefont
  {Baruchel}, \citenamefont {Schlenker},\ and\ \citenamefont
  {Roth}}]{Baruchel:1977hy}%
  \BibitemOpen
  \bibfield  {author} {\bibinfo {author} {\bibfnamefont {J.}~\bibnamefont
  {Baruchel}}, \bibinfo {author} {\bibfnamefont {M.}~\bibnamefont {Schlenker}},
  \ and\ \bibinfo {author} {\bibfnamefont {W.~L.}\ \bibnamefont {Roth}},\
  }\href {\doibase 10.1063/1.323361} {\bibfield  {journal} {\bibinfo  {journal}
  {Journal of Applied Physics}\ }\textbf {\bibinfo {volume} {48}},\ \bibinfo
  {pages} {5} (\bibinfo {year} {1977})}\BibitemShut {NoStop}%
\bibitem [{\citenamefont {Arai}\ \emph {et~al.}(2012)\citenamefont {Arai},
  \citenamefont {Okuda}, \citenamefont {Tanaka}, \citenamefont {Kotsugi},
  \citenamefont {Fukumoto}, \citenamefont {Ohkochi}, \citenamefont {Nakamura},
  \citenamefont {Matsushita}, \citenamefont {Muro}, \citenamefont {Oura},
  \citenamefont {Senba}, \citenamefont {Ohashi}, \citenamefont {Kakizaki},
  \citenamefont {Mitsumata},\ and\ \citenamefont {Kinoshita}}]{Arai2012}%
  \BibitemOpen
  \bibfield  {author} {\bibinfo {author} {\bibfnamefont {K.}~\bibnamefont
  {Arai}}, \bibinfo {author} {\bibfnamefont {T.}~\bibnamefont {Okuda}},
  \bibinfo {author} {\bibfnamefont {A.}~\bibnamefont {Tanaka}}, \bibinfo
  {author} {\bibfnamefont {M.}~\bibnamefont {Kotsugi}}, \bibinfo {author}
  {\bibfnamefont {K.}~\bibnamefont {Fukumoto}}, \bibinfo {author}
  {\bibfnamefont {T.}~\bibnamefont {Ohkochi}}, \bibinfo {author} {\bibfnamefont
  {T.}~\bibnamefont {Nakamura}}, \bibinfo {author} {\bibfnamefont
  {T.}~\bibnamefont {Matsushita}}, \bibinfo {author} {\bibfnamefont
  {T.}~\bibnamefont {Muro}}, \bibinfo {author} {\bibfnamefont {M.}~\bibnamefont
  {Oura}}, \bibinfo {author} {\bibfnamefont {Y.}~\bibnamefont {Senba}},
  \bibinfo {author} {\bibfnamefont {H.}~\bibnamefont {Ohashi}}, \bibinfo
  {author} {\bibfnamefont {A.}~\bibnamefont {Kakizaki}}, \bibinfo {author}
  {\bibfnamefont {C.}~\bibnamefont {Mitsumata}}, \ and\ \bibinfo {author}
  {\bibfnamefont {T.}~\bibnamefont {Kinoshita}},\ }\href {\doibase
  10.1103/PhysRevB.85.104418} {\bibfield  {journal} {\bibinfo  {journal}
  {Physical Review B}\ }\textbf {\bibinfo {volume} {85}},\ \bibinfo {pages}
  {104418} (\bibinfo {year} {2012})}\BibitemShut {NoStop}%
\bibitem [{\citenamefont {Ozhogin}(1976)}]{Ozhogin:1976hu}%
  \BibitemOpen
  \bibfield  {author} {\bibinfo {author} {\bibfnamefont {V.}~\bibnamefont
  {Ozhogin}},\ }\href {\doibase 10.1109/TMAG.1976.1058992} {\bibfield
  {journal} {\bibinfo  {journal} {IEEE Transactions on Magnetics}\ }\textbf
  {\bibinfo {volume} {12}},\ \bibinfo {pages} {19} (\bibinfo {year}
  {1976})}\BibitemShut {NoStop}%
\bibitem [{SM()}]{SM}%
  \BibitemOpen
  \href@noop {} {}\bibinfo {note} {See Supplemental Material for (i) Static
  absorption spectroscopy and exciton-magnon mechanism (ii) Experimental
  pump-probe set-up (iii) Pump-probe data analysis (IV) Probe-polarization
  dependence (V) Pump-probe data in a single-domain (VI) Theoretical model,
  which includes the following references: [13], [26], [27],
  [37]-[45]}\BibitemShut {NoStop}%
\bibitem [{\citenamefont {Grimsditch}\ \emph {et~al.}(1998)\citenamefont
  {Grimsditch}, \citenamefont {McNeil},\ and\ \citenamefont
  {Lockwood}}]{grimsditch1998}%
  \BibitemOpen
  \bibfield  {author} {\bibinfo {author} {\bibfnamefont {M.}~\bibnamefont
  {Grimsditch}}, \bibinfo {author} {\bibfnamefont {L.~E.}\ \bibnamefont
  {McNeil}}, \ and\ \bibinfo {author} {\bibfnamefont {D.~J.}\ \bibnamefont
  {Lockwood}},\ }\href {\doibase 10.1103/physrevb.58.14462} {\bibfield
  {journal} {\bibinfo  {journal} {Physical Review B}\ }\textbf {\bibinfo
  {volume} {58}},\ \bibinfo {pages} {14462} (\bibinfo {year}
  {1998})}\BibitemShut {NoStop}%
\bibitem [{\citenamefont {Takahara}\ \emph {et~al.}(2012)\citenamefont
  {Takahara}, \citenamefont {Jinn}, \citenamefont {Wakabayashi}, \citenamefont
  {Moriyasu},\ and\ \citenamefont {Kohmoto}}]{Takahara:2012cp}%
  \BibitemOpen
  \bibfield  {author} {\bibinfo {author} {\bibfnamefont {M.}~\bibnamefont
  {Takahara}}, \bibinfo {author} {\bibfnamefont {H.}~\bibnamefont {Jinn}},
  \bibinfo {author} {\bibfnamefont {S.}~\bibnamefont {Wakabayashi}}, \bibinfo
  {author} {\bibfnamefont {T.}~\bibnamefont {Moriyasu}}, \ and\ \bibinfo
  {author} {\bibfnamefont {T.}~\bibnamefont {Kohmoto}},\ }\href {\doibase
  10.1103/PhysRevB.86.094301} {\bibfield  {journal} {\bibinfo  {journal}
  {Physical Review B}\ }\textbf {\bibinfo {volume} {86}},\ \bibinfo {pages}
  {094301} (\bibinfo {year} {2012})}\BibitemShut {NoStop}%
\bibitem [{\citenamefont {Tanabe}\ \emph {et~al.}(1965)\citenamefont {Tanabe},
  \citenamefont {Moriya},\ and\ \citenamefont {Sugano}}]{Tanabe1965}%
  \BibitemOpen
  \bibfield  {author} {\bibinfo {author} {\bibfnamefont {Y.}~\bibnamefont
  {Tanabe}}, \bibinfo {author} {\bibfnamefont {T.}~\bibnamefont {Moriya}}, \
  and\ \bibinfo {author} {\bibfnamefont {S.}~\bibnamefont {Sugano}},\ }\href
  {\doibase 10.1103/PhysRevLett.15.1023} {\bibfield  {journal} {\bibinfo
  {journal} {Physical Review Letters}\ }\textbf {\bibinfo {volume} {15}},\
  \bibinfo {pages} {1023} (\bibinfo {year} {1965})}\BibitemShut {NoStop}%
\bibitem [{\citenamefont {Tanabe}\ and\ \citenamefont
  {Aoyagi}(1982)}]{Tanabe:1982vb}%
  \BibitemOpen
  \bibfield  {author} {\bibinfo {author} {\bibfnamefont {Y.}~\bibnamefont
  {Tanabe}}\ and\ \bibinfo {author} {\bibfnamefont {K.}~\bibnamefont
  {Aoyagi}},\ }\href@noop {} {{{0 444 86202 1}\emph {\bibinfo
  {title} {{Excitons in magnetic insulators}}}}},\ edited by\ \bibinfo {editor}
  {\bibfnamefont {E.~I.}\ \bibnamefont {Rashba}}\ and\ \bibinfo {editor}
  {\bibfnamefont {M.~D.}\ \bibnamefont {Sturge}}\ (\bibinfo  {publisher}
  {North-Holland},\ \bibinfo {year} {1982})\BibitemShut {NoStop}%
\bibitem [{\citenamefont {Mironova-Ulmane}\ \emph {et~al.}(2012)\citenamefont
  {Mironova-Ulmane}, \citenamefont {Skvortsova}, \citenamefont {Kuzmin},\ and\
  \citenamefont {Sildos}}]{MironovaUlmane:2012ca}%
  \BibitemOpen
  \bibfield  {author} {\bibinfo {author} {\bibfnamefont {N.}~\bibnamefont
  {Mironova-Ulmane}}, \bibinfo {author} {\bibfnamefont {V.}~\bibnamefont
  {Skvortsova}}, \bibinfo {author} {\bibfnamefont {A.}~\bibnamefont {Kuzmin}},
  \ and\ \bibinfo {author} {\bibfnamefont {I.}~\bibnamefont {Sildos}},\ }in\
  \href {\doibase 10.1117/12.639158} {\emph {\bibinfo {booktitle} {SPIE
  Proceedings}}},\ \bibinfo {editor} {edited by\ \bibinfo {editor}
  {\bibfnamefont {A.}~\bibnamefont {Rosental}}}\ (\bibinfo  {publisher}
  {SPIE},\ \bibinfo {year} {2012})\ pp.\ \bibinfo {pages}
  {59460D--59460D--5}\BibitemShut {NoStop}%
\bibitem [{\citenamefont {Bossini}\ \emph {et~al.}(2014)\citenamefont
  {Bossini}, \citenamefont {Kalashnikova}, \citenamefont {Pisarev},
  \citenamefont {Rasing},\ and\ \citenamefont {Kimel}}]{Bossini2014}%
  \BibitemOpen
  \bibfield  {author} {\bibinfo {author} {\bibfnamefont {D.}~\bibnamefont
  {Bossini}}, \bibinfo {author} {\bibfnamefont {A.~M.}\ \bibnamefont
  {Kalashnikova}}, \bibinfo {author} {\bibfnamefont {R.~V.}\ \bibnamefont
  {Pisarev}}, \bibinfo {author} {\bibfnamefont {{\relax Th}.}~\bibnamefont
  {Rasing}}, \ and\ \bibinfo {author} {\bibfnamefont {A.~V.}\ \bibnamefont
  {Kimel}},\ }\href {\doibase 10.1103/PhysRevB.89.060405} {\bibfield  {journal}
  {\bibinfo  {journal} {Physical Review B}\ }\textbf {\bibinfo {volume} {89}},\
  \bibinfo {pages} {060405(R)} (\bibinfo {year} {2014})}\BibitemShut {NoStop}%
\bibitem [{\citenamefont {Kimel}\ \emph {et~al.}(2002)\citenamefont {Kimel},
  \citenamefont {Pisarev}, \citenamefont {Hohlfeld},\ and\ \citenamefont
  {Rasing}}]{Kimel2002}%
  \BibitemOpen
  \bibfield  {author} {\bibinfo {author} {\bibfnamefont {A.~V.}\ \bibnamefont
  {Kimel}}, \bibinfo {author} {\bibfnamefont {R.~V.}\ \bibnamefont {Pisarev}},
  \bibinfo {author} {\bibfnamefont {J.}~\bibnamefont {Hohlfeld}}, \ and\
  \bibinfo {author} {\bibfnamefont {{\relax Th}.}~\bibnamefont {Rasing}},\
  }\href {\doibase 10.1103/physrevlett.89.287401} {\bibfield  {journal}
  {\bibinfo  {journal} {Phys. Rev. Lett.}\ }\textbf {\bibinfo {volume} {89}},\
  \bibinfo {pages} {287401} (\bibinfo {year} {2002})}\BibitemShut {NoStop}%
\bibitem [{\citenamefont {Ganot}\ \emph {et~al.}(1982)\citenamefont {Ganot},
  \citenamefont {Dugautier}, \citenamefont {Moch},\ and\ \citenamefont
  {Nouet}}]{Ganot:1982bj}%
  \BibitemOpen
  \bibfield  {author} {\bibinfo {author} {\bibfnamefont {F.}~\bibnamefont
  {Ganot}}, \bibinfo {author} {\bibfnamefont {C.}~\bibnamefont {Dugautier}},
  \bibinfo {author} {\bibfnamefont {P.}~\bibnamefont {Moch}}, \ and\ \bibinfo
  {author} {\bibfnamefont {J.}~\bibnamefont {Nouet}},\ }\href {\doibase
  10.1088/0022-3719/15/4/026} {\bibfield  {journal} {\bibinfo  {journal}
  {Journal of Physics C: Solid State Physics}\ }\textbf {\bibinfo {volume}
  {15}},\ \bibinfo {pages} {801} (\bibinfo {year} {1982})}\BibitemShut
  {NoStop}%
\bibitem [{\citenamefont {Bossini}\ \emph {et~al.}(2018)\citenamefont
  {Bossini}, \citenamefont {Konishi}, \citenamefont {Toyoda}, \citenamefont
  {Arima}, \citenamefont {Yumoto},\ and\ \citenamefont
  {Kuwata-Gonokami}}]{Bossini:2018db}%
  \BibitemOpen
  \bibfield  {author} {\bibinfo {author} {\bibfnamefont {D.}~\bibnamefont
  {Bossini}}, \bibinfo {author} {\bibfnamefont {K.}~\bibnamefont {Konishi}},
  \bibinfo {author} {\bibfnamefont {S.}~\bibnamefont {Toyoda}}, \bibinfo
  {author} {\bibfnamefont {T.}~\bibnamefont {Arima}}, \bibinfo {author}
  {\bibfnamefont {J.}~\bibnamefont {Yumoto}}, \ and\ \bibinfo {author}
  {\bibfnamefont {M.}~\bibnamefont {Kuwata-Gonokami}},\ }\href {\doibase
  10.1038/s41567-017-0036-1} {\bibfield  {journal} {\bibinfo  {journal} {Nature
  Physics}\ }\textbf {\bibinfo {volume} {14}},\ \bibinfo {pages} {370}
  (\bibinfo {year} {2018})}\BibitemShut {NoStop}%
\bibitem [{\citenamefont {Kimel}\ \emph {et~al.}(2005)\citenamefont {Kimel},
  \citenamefont {Kirilyuk}, \citenamefont {Usachev}, \citenamefont {Pisarev},
  \citenamefont {Balbashov},\ and\ \citenamefont {Rasing}}]{Kimel2005}%
  \BibitemOpen
  \bibfield  {author} {\bibinfo {author} {\bibfnamefont {A.~V.}\ \bibnamefont
  {Kimel}}, \bibinfo {author} {\bibfnamefont {A.}~\bibnamefont {Kirilyuk}},
  \bibinfo {author} {\bibfnamefont {P.~A.}\ \bibnamefont {Usachev}}, \bibinfo
  {author} {\bibfnamefont {R.~V.}\ \bibnamefont {Pisarev}}, \bibinfo {author}
  {\bibfnamefont {A.~M.}\ \bibnamefont {Balbashov}}, \ and\ \bibinfo {author}
  {\bibfnamefont {{\relax Th}.}~\bibnamefont {Rasing}},\ }\href {\doibase Doi
  10.1038/Nature03564} {\bibfield  {journal} {\bibinfo  {journal} {Nature}\
  }\textbf {\bibinfo {volume} {435}},\ \bibinfo {pages} {655} (\bibinfo {year}
  {2005})}\BibitemShut {NoStop}%
\bibitem [{\citenamefont {Kirilyuk}\ \emph {et~al.}(2010)\citenamefont
  {Kirilyuk}, \citenamefont {Kimel},\ and\ \citenamefont
  {Rasing}}]{Kirilyuk2010}%
  \BibitemOpen
  \bibfield  {author} {\bibinfo {author} {\bibfnamefont {A.}~\bibnamefont
  {Kirilyuk}}, \bibinfo {author} {\bibfnamefont {A.~V.}\ \bibnamefont {Kimel}},
  \ and\ \bibinfo {author} {\bibfnamefont {{\relax Th}.}~\bibnamefont
  {Rasing}},\ }\href {\doibase 10.1103/RevModPhys.82.2731} {\bibfield
  {journal} {\bibinfo  {journal} {Reviews of Modern Physics}\ }\textbf
  {\bibinfo {volume} {82}},\ \bibinfo {pages} {2731} (\bibinfo {year}
  {2010})}\BibitemShut {NoStop}%
\bibitem [{\citenamefont {Bar'yakhtar}\ and\ \citenamefont
  {Ivanov}(1980)}]{BarYakhtar:1980dz}%
  \BibitemOpen
  \bibfield  {author} {\bibinfo {author} {\bibfnamefont {I.~V.}\ \bibnamefont
  {Bar'yakhtar}}\ and\ \bibinfo {author} {\bibfnamefont {B.~A.}\ \bibnamefont
  {Ivanov}},\ }\href {\doibase 10.1016/0038-1098(80)90148-9} {\bibfield
  {journal} {\bibinfo  {journal} {Solid State Communications}\ }\textbf
  {\bibinfo {volume} {34}},\ \bibinfo {pages} {545} (\bibinfo {year}
  {1980})}\BibitemShut {NoStop}%
\bibitem [{\citenamefont {Bossini}\ and\ \citenamefont
  {Rasing}(2017)}]{Bossini:2017bo}%
  \BibitemOpen
  \bibfield  {author} {\bibinfo {author} {\bibfnamefont {D.}~\bibnamefont
  {Bossini}}\ and\ \bibinfo {author} {\bibfnamefont {{\relax Th}.}~\bibnamefont
  {Rasing}},\ }\href {\doibase 10.1088/1402-4896/aa54d4} {\bibfield  {journal}
  {\bibinfo  {journal} {Physica Scripta}\ }\textbf {\bibinfo {volume} {92}},\
  \bibinfo {pages} {024002} (\bibinfo {year} {2017})}\BibitemShut {NoStop}%
\bibitem [{\citenamefont {Gomonay}\ and\ \citenamefont
  {Bossini}(2021)}]{Gomonay2021}%
  \BibitemOpen
  \bibfield  {author} {\bibinfo {author} {\bibfnamefont {O.}~\bibnamefont
  {Gomonay}}\ and\ \bibinfo {author} {\bibfnamefont {D.}~\bibnamefont
  {Bossini}},\ }\href@noop {} {\bibfield  {journal} {\bibinfo  {journal}
  {Journal of Physics D: Applied Physics}\ }\newline
  \bibinfo {pages} {http://iopscience.iop.org/article/10.1088/1361-6463/ac055c} (\bibinfo {year}
  {2021})}\BibitemShut {NoStop}%
\bibitem [{\citenamefont {Reichardt}\ \emph {et~al.}(1975)\citenamefont
  {Reichardt}, \citenamefont {Wagner},\ and\ \citenamefont
  {Kress}}]{Reichardt:1975bm}%
  \BibitemOpen
  \bibfield  {author} {\bibinfo {author} {\bibfnamefont {W.}~\bibnamefont
  {Reichardt}}, \bibinfo {author} {\bibfnamefont {V.}~\bibnamefont {Wagner}}, \
  and\ \bibinfo {author} {\bibfnamefont {W.}~\bibnamefont {Kress}},\ }\href
  {\doibase 10.1088/0022-3719/8/23/009} {\bibfield  {journal} {\bibinfo
  {journal} {Journal of Physics C: Solid State Physics}\ }\textbf {\bibinfo
  {volume} {8}},\ \bibinfo {pages} {3955 } (\bibinfo {year}
  {1975})}\BibitemShut {NoStop}%
\bibitem [{\citenamefont {Stancik}\ and\ \citenamefont
  {Brauns}(2008)}]{Stancik2008}%
  \BibitemOpen
  \bibfield  {author} {\bibinfo {author} {\bibfnamefont {A.~L.}\ \bibnamefont
  {Stancik}}\ and\ \bibinfo {author} {\bibfnamefont {E.~B.}\ \bibnamefont
  {Brauns}},\ }\href {\doibase 10.1016/j.vibspec.2008.02.009} {\bibfield
  {journal} {\bibinfo  {journal} {Vibrational Spectroscopy}\ }\textbf {\bibinfo
  {volume} {47}},\ \bibinfo {pages} {66 } (\bibinfo {year} {2008})}\BibitemShut
  {NoStop}%
\bibitem [{\citenamefont {Sell}(1968)}]{Sell1968}%
  \BibitemOpen
  \bibfield  {author} {\bibinfo {author} {\bibfnamefont {D.~D.}\ \bibnamefont
  {Sell}},\ }\href {\doibase 10.1063/1.1656158} {\bibfield  {journal} {\bibinfo
   {journal} {Journal of Applied Physics}\ }\textbf {\bibinfo {volume} {39}},\
  \bibinfo {pages} {1030} (\bibinfo {year} {1968})}\BibitemShut {NoStop}%
\bibitem [{\citenamefont {Sell}\ \emph {et~al.}(1967)\citenamefont {Sell},
  \citenamefont {Greene},\ and\ \citenamefont {White}}]{Sell:1967jo}%
  \BibitemOpen
  \bibfield  {author} {\bibinfo {author} {\bibfnamefont {D.~D.}\ \bibnamefont
  {Sell}}, \bibinfo {author} {\bibfnamefont {R.~L.}\ \bibnamefont {Greene}}, \
  and\ \bibinfo {author} {\bibfnamefont {R.~M.}\ \bibnamefont {White}},\ }\href
  {\doibase 10.1103/physrev.158.489} {\bibfield  {journal} {\bibinfo  {journal}
  {Physical Review}\ }\textbf {\bibinfo {volume} {158}},\ \bibinfo {pages} {489
  } (\bibinfo {year} {1967})}\BibitemShut {NoStop}%
\bibitem [{\citenamefont {Macfarlane}\ and\ \citenamefont
  {Allen}(1971)}]{Macfarlane1971}%
  \BibitemOpen
  \bibfield  {author} {\bibinfo {author} {\bibfnamefont {R.~M.}\ \bibnamefont
  {Macfarlane}}\ and\ \bibinfo {author} {\bibfnamefont {J.~W.}\ \bibnamefont
  {Allen}},\ }\href {\doibase 10.1103/physrevb.4.3054} {\bibfield  {journal}
  {\bibinfo  {journal} {Physical Review B}\ }\textbf {\bibinfo {volume} {4}},\
  \bibinfo {pages} {3054 } (\bibinfo {year} {1971})}\BibitemShut {NoStop}%
\bibitem [{\citenamefont {Polley}\ \emph {et~al.}(2018)\citenamefont {Polley},
  \citenamefont {Pancaldi}, \citenamefont {Hudl}, \citenamefont {Vavassori},
  \citenamefont {Urazhdin},\ and\ \citenamefont {Bonetti}}]{Polley:2018cm}%
  \BibitemOpen
  \bibfield  {author} {\bibinfo {author} {\bibfnamefont {D.}~\bibnamefont
  {Polley}}, \bibinfo {author} {\bibfnamefont {M.}~\bibnamefont {Pancaldi}},
  \bibinfo {author} {\bibfnamefont {M.}~\bibnamefont {Hudl}}, \bibinfo {author}
  {\bibfnamefont {P.}~\bibnamefont {Vavassori}}, \bibinfo {author}
  {\bibfnamefont {S.}~\bibnamefont {Urazhdin}}, \ and\ \bibinfo {author}
  {\bibfnamefont {S.}~\bibnamefont {Bonetti}},\ }\href {\doibase
  10.1088/1361-6463/aaa863} {\bibfield  {journal} {\bibinfo  {journal} {Journal
  of Physics D: Applied Physics}\ }\textbf {\bibinfo {volume} {51}},\ \bibinfo
  {pages} {084001} (\bibinfo {year} {2018})}\BibitemShut {NoStop}%
\bibitem [{\citenamefont {Satoh}\ \emph
  {et~al.}(2010{\natexlab{b}})\citenamefont {Satoh}, \citenamefont {Cho},
  \citenamefont {Shimura}, \citenamefont {Kuroda}, \citenamefont {Ueda},
  \citenamefont {Ueda},\ and\ \citenamefont {Fiebig}}]{Satoh2010Japp}%
  \BibitemOpen
  \bibfield  {author} {\bibinfo {author} {\bibfnamefont {T.}~\bibnamefont
  {Satoh}}, \bibinfo {author} {\bibfnamefont {S.-J.}\ \bibnamefont {Cho}},
  \bibinfo {author} {\bibfnamefont {T.}~\bibnamefont {Shimura}}, \bibinfo
  {author} {\bibfnamefont {K.}~\bibnamefont {Kuroda}}, \bibinfo {author}
  {\bibfnamefont {H.}~\bibnamefont {Ueda}}, \bibinfo {author} {\bibfnamefont
  {Y.}~\bibnamefont {Ueda}}, \ and\ \bibinfo {author} {\bibfnamefont
  {M.}~\bibnamefont {Fiebig}},\ }\href {\doibase 10.1364/josab.27.001421}
  {\bibfield  {journal} {\bibinfo  {journal} {Journal of the Optical Society of
  America B}\ }\textbf {\bibinfo {volume} {27}},\ \bibinfo {pages} {1421}
  (\bibinfo {year} {2010}{\natexlab{b}})}\BibitemShut {NoStop}%
\bibitem [{\citenamefont {Herrmann-Ronzaud}\ \emph {et~al.}(1978)\citenamefont
  {Herrmann-Ronzaud}, \citenamefont {Burlet},\ and\ \citenamefont
  {Rossat-Mignod}}]{Herrmann_Ronzaud_1978}%
  \BibitemOpen
  \bibfield  {author} {\bibinfo {author} {\bibfnamefont {D.}~\bibnamefont
  {Herrmann-Ronzaud}}, \bibinfo {author} {\bibfnamefont {P.}~\bibnamefont
  {Burlet}}, \ and\ \bibinfo {author} {\bibfnamefont {J.}~\bibnamefont
  {Rossat-Mignod}},\ }\href {\doibase 10.1088/0022-3719/11/10/023} {\bibfield
  {journal} {\bibinfo  {journal} {Journal of Physics C: Solid State Physics}\
  }\textbf {\bibinfo {volume} {11}},\ \bibinfo {pages} {2123} (\bibinfo {year}
  {1978})}\BibitemShut {NoStop}%
\bibitem [{\citenamefont {P\"{o}schel}\ and\ \citenamefont
  {Teller}(1933)}]{Poeschel1933}%
  \BibitemOpen
  \bibfield  {author} {\bibinfo {author} {\bibfnamefont {G.}~\bibnamefont
  {P\"{o}schel}}\ and\ \bibinfo {author} {\bibfnamefont {E.}~\bibnamefont
  {Teller}},\ }\href@noop {} {\bibfield  {journal} {\bibinfo  {journal}
  {Zeitschrift f{\"u}r Physik}\ }\textbf {\bibinfo {volume} {83}},\ \bibinfo
  {pages} {143} (\bibinfo {year} {1933})}\BibitemShut {NoStop}%
\bibitem [{\citenamefont {Cooper}\ \emph {et~al.}(1995)\citenamefont {Cooper},
  \citenamefont {Khare},\ and\ \citenamefont {Sukhatme}}]{Cooper1995}%
  \BibitemOpen
  \bibfield  {author} {\bibinfo {author} {\bibfnamefont {F.}~\bibnamefont
  {Cooper}}, \bibinfo {author} {\bibfnamefont {A.}~\bibnamefont {Khare}}, \
  and\ \bibinfo {author} {\bibfnamefont {U.}~\bibnamefont {Sukhatme}},\ }\href
  {\doibase https://doi.org/10.1016/0370-1573(94)00080-M} {\bibfield  {journal}
  {\bibinfo  {journal} {Physics Reports}\ }\textbf {\bibinfo {volume} {251}},\
  \bibinfo {pages} {267} (\bibinfo {year} {1995})}\BibitemShut {NoStop}%
\end{thebibliography}
%

\end{document}